\documentclass{amsart}
\usepackage{amsmath}
\usepackage{graphicx}
\usepackage{booktabs}
\title[Simultaneous estimation of $R$ and the detection rate]{Simultaneous estimation of the effective reproducing number and the detection rate of COVID-19}
\author{Yoriyuki Yamagata}
\address{National Institute of Advanced Industrial Science and Technology (AIST),
1-8-31 Midorigaoka, Ikeda, Japan}
\email{yoriyuki.yamagata@aist.go.jp}
\date{\today}

\begin{document}

\maketitle

\begin{abstract}
 A major difficulty to estimate $R$ (the effective reproducing number) of COVID-19 is that most cases of COVID-19 infection are mild or asymptomatic, therefore true number of infection is difficult to determine.
 This paper estimates the daily change of $R$ and the detection rate simultaneously using a Bayesian model.
 The analysis using synthesized data shows that our model correctly estimates $R$ and detects a short-term shock of the detection rate.
 Then, we apply our model to data from several countries to evaluate the effectiveness of public healthcare measures.
 Our analysis focuses Japan, which employs a moderate measure to keep ``social distance''.
 The result indicates a downward trend and now $R$ becomes below $1$.
 Although our analysis is preliminary, this may suggest that a moderate policy still can prevent epidemic of COVID-19.
\end{abstract}

\section{Introduction}

In the wake of the COVID-19 epidemic, the Japanese government gradually employed public health measures against COVID-19.
In the first stage, the stronger quarantine measures at the border were implemented.
Once patients who had no connection to Wuhan appeared inside the border, The government started to track these patients as far as possible and tried to find people contacted to these patients.
Once these ``track and quarantine clusters'' tactics were overwhelmed by the number of patients, the government started to ask behavior changes to people, culminating ``declaration of the emergency'' on April 6.
Public facilities like libraries were closed.
Large shopping malls and entertainment businesses, such as movie theatres, were asked to be closed.
Restaurants are asked to shorten their operating hours and stop providing alcoholic beverages at night.
Working from home was encouraged, and the citizens were advised to avoid crowded areas and generally avoid going outside unnecessarily.
However, the Japanese legal system does not have enough mechanisms to enforce these policies.
Thus the effectiveness of these policies can be questioned.
Many people are still commuting to their office because many companies lack the necessary ability to allow their employees to work from home.
Many small restaurants and cafes are still running their business because of the lack of financial compensation.

To estimate the policy effect to COVID-19, we need estimate daily changes of the effective reproducing number $R$.
A major difficulty to estimate $R$ of COVID-19 is that most cases of COVID-19 infection are mild or asymptomatic, therefore the true number of infection is difficult to determine.
This paper estimates the daily change of $R$ and the detection rate simultaneously using a Bayesian model.
The analysis using synthesized data shows that our model correctly estimates $R$ and detects a short-term shock of the detection rate.
Our analysis focuses Japan, which employs a moderate measure to keep ``social distance''.
The result indicates a downward trend and now $R$ becomes below $1$.
Although our analysis is preliminary, this may suggest that a moderate policy still can prevent epidemic of COVID-19.

Then, we apply our model to data from several countries to evaluate the effectiveness of public healthcare measures.
The comparison between Denmark and Sweden reveals that lock-down is very effective in short term.
However, In Sweden, which did not employ lock-down, $R$ also reduced and now $R$ of both countries are roughly same.
This might suggest that lock-down is not effective long run, or Denmark is disadvantageous against COVID-19 comparing to Sweden.

Further, we applied our method to China, Italy and US and show that these countries are also about exiting from epidemic.

\section{Related works}

Several works employ data-driven methods to predict and measure the public health measure of COVID-19.

Anastassopoulou et al.~\cite{Anastassopoulou2020} apply the SIRD model to Chinese official statistics, estimating parameters using linear regression.
The reporting rate is not estimated from data but assumed.
By these models and parameters, they predict the COVID-19 epidemic in Hubei province.

Diego Caccavo~\cite{Caccavo2020} and independently Peter Turchin~\cite{Turchin2020} apply modified SIRD models, in which parameters change overtime following specific function forms.
Parameters govern these functions are estimated by minimizing the sum-of-square-error.
However, using the sum-of-square method causes over-fitting and always favors a complex model, therefore it is not suitable to access policy effectiveness.
Further, fitting the SIRD model in the early stage of infection is difficult, as pointed out in stat-exchange~\footnote{https://stats.stackexchange.com/questions/446712/fitting-sir-model-with-2019-ncov-data-doesnt-conververge}.
Using a Bayesian method, we avoid these problems to some degree, because a Bayesian method estimates parameter distribution instead of a point estimate.
Thus, we can assess the degree of confidence of each parameter.
Further, by well-established statistical methods, we can compare the explanatory power of different models.

Flaxman et al.~\cite{Flaxman2020} use a Bayesian model to estimate policy effectiveness.
The methodology is different from us because they assume immediate effects from the policies implemented.
Further, they use a discrete renewal process, a more advanced model than the SIRD model.
They use parameters estimated from studies of clinical cases while we use a purely data-driven method.

\section{Method and materials}

\subsection{Model}

We use the discrete-time SIR model but assume that the number of move between each category is stochastic and follows Poisson distribution.

\begin{align}
 NI(t) &\sim \textup{Poisson}(\frac{\beta I S}{P})\\
 NR(t) &\sim \textup{Poisson}(\gamma I)\\
 I(t+1) &= I(t) + NI(t) - NR(t)\\
 S(t+1) &= S(t) - NI(t)\\
 R(t+1) &= R(t) + NR(t)
\end{align}
The effective reproduction rate can be written
\begin{equation}
 R = \frac{\beta}{\gamma} \cdot \frac{S}{P}
\end{equation}
When $R$ becomes $<1$ then the infection starts to decline.

We cannot expect that these values are directly observable, because many (or most) cases are mild or asymptomatic.
Therefore, we introduce the detection rate $q$ and let the number of cumulative observed cases $C_{\text{obs}}$ and recovered $R_{\text{obs}}$ as
\begin{align}
 NI_{\text{obs}}(t) &\sim \textup{Poisson}(q NI(t))\\
 NR_{\text{obs}}(t) &\sim \textup{Poisson}(\gamma I_{\text{obs}}(t))\\
 I_{\text{obs}}(t+1) &= I_{\text{obs}}(t) + NI_{\text{obs}}(t) - NR_{\text{obs}}(t)\\
 R_{\text{obs}}(t+1) &= R_{\text{obs}}(t) + NR_{\text{obs}}(t)
\end{align}

We assume that $\beta$ and $q$ change day to day bases while other parameters are fixed.
To get a reasonable estimate, we assume prior distributions somewhat arbitrary chosen.
\begin{gather}
 S(0) = I(0) \sim \textup{Student\_t}(3, 0, 1)\\
 \beta(0) \sim \textup{Student\_t}(3, 0, 1)\\
 \beta(t+1) \sim \textup{Student\_t}(3, \beta(t), \sigma_b)\\
 \sigma_b \sim \textup{Exponential}(1)
\end{gather}
while the prior of $q(t)$ is the uniform distribution over $[0, 1]$.
To make our model robust, we choose Student-t as the prior for $\beta$.
We perform a sensitivity analysis of the prior for $\sigma_b$ to ensure that the choice of priors does not strongly affect the result.



\subsection{Data}

The number of confirmed cases of each country up to May 11 were drawn from data repository~\footnote{https://github.com/CSSEGISandData/COVID-19} by Johns Hopkins University Center for Systems Science and Engineering.
The dataset also contains the number of recovery, but as pointed out in README the number is underestimated.
For example, the number of recovery in Norway is only 32 in May 11, which is impossible with 8132 confirmed cases.
Therefore, we estimate $\gamma = 0.04$ per day by Chinese data and assume that $\gamma$ is constant across all countries.

\subsection{Experiment}

First, we performed model validation using synthesized data of the scenarios in which $\beta$ and $q$ are constant, $\beta$ is constant but $q$ is piecewise constant, $\beta$ reduces linearly and $q$ is constant with white noise and $\beta, q$ are constant where $\beta$ is high enough to make almost all people obtain immunity.
The result is presented in Section~\ref{sec:validation}.

Then the real-world data were fed to Stan~\cite{carpenter2017stan} for parameter estimation by our Bayesian model.
We simplified our model to ease modeling in Stan.
Because latent discrete variables cannot be used in Stan, we used real numbers for $NI(t)$ and $NR(t)$.
We used normal approximation $\mathcal{N}(\lambda, \sqrt{\lambda})$ for the Poisson distribution used for $NI(t)$.
For $NR(t)$, we replaced stochastic laws to deterministic laws $NR(t) = \gamma I$ to avoid a numerical issue.

Parameter estimation used 10,000 iterations with 5,000 iterations for warm-up and 5,000 iterations for sampling.
Four (default number of Stan) independent computations were performed simultaneously and used to estimate $\hat{R}$.
If $\hat{R} < 1.1$ then we regard that the estimation is converged.

To make sure that our results are meaningful, we compared the performance of our model with a model (we refer it by ``the constant model'' \textbf{const}), which assumes constant $\beta$ and $q$ and a model (we refer it by ``the constant detection rate model'' \textbf{const-q}).
\textbf{const} estimates constant $\beta$ and $q$ simultaneously but \textbf{const-q} takes $q$ as a part of given data.
Parameters were estimated for \textbf{const} and \textbf{const-q} in the same way and LOO, a standard measure of model performance, was compared.
Because the exact computation of LOO-CV is computationally expensive, approximation PSIS-LOO-CV~\cite{Vehtari2017} and WAIC~\cite{watanabe2010asymptotic} were compared.
Further, we check the reliability of these estimation by checking the Pareto-k for the importance weight distribution.
The computation of PSIS-LOO-CV and WAIC is performed by ArviZ~\cite{arviz2019}.

Models and the computation history used for this experiment are public at GitHub~\footnote{https://github.com/yoriyuki/BayesianCOVID19/tree/paper-version-3/notebook}.

\section{Results}

\subsection{Model validation}\label{sec:validation}

\begin{figure}[h]
 \centering
 \includegraphics[width=\linewidth]{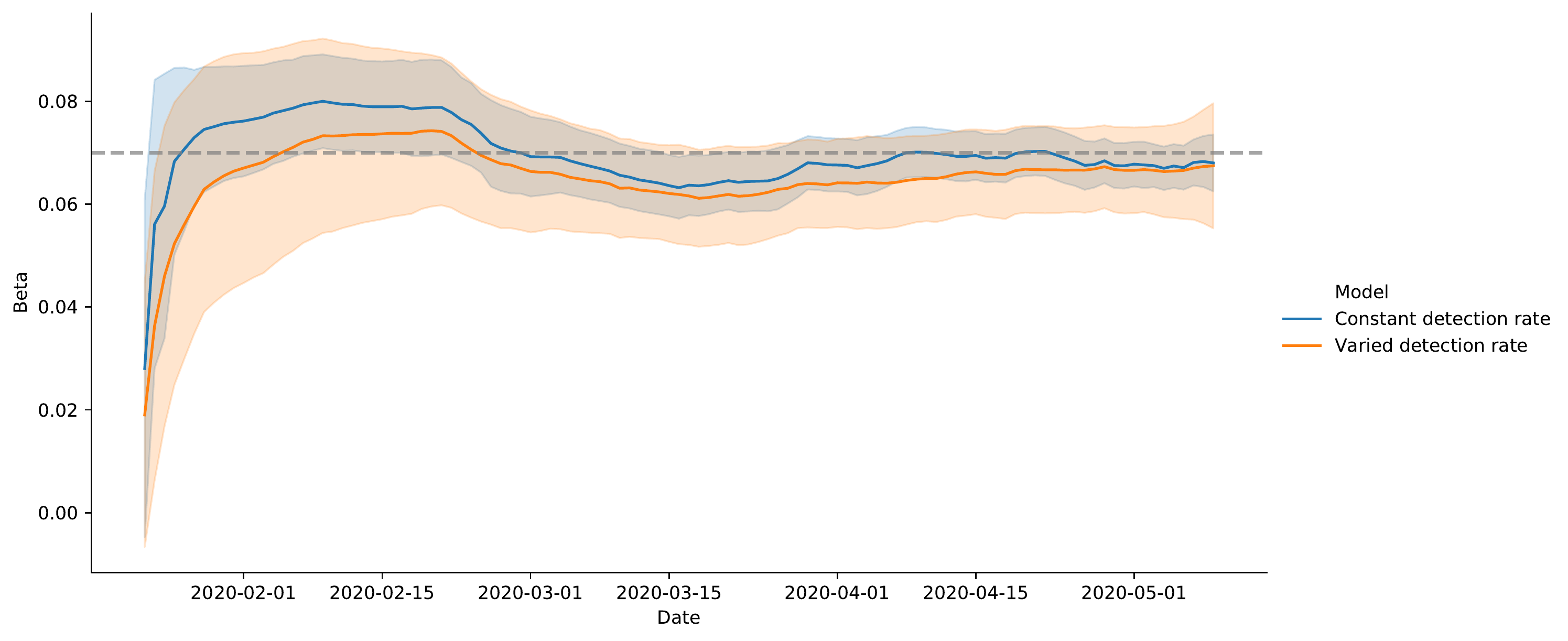}
 \caption{Estimated $\beta$ for data generated with constant $\beta$ and $q$. The blue line indicates the estimate of \textbf{const-q} model while the orange line indicates the estimate of our model. The dotted line indicates the true $\beta$. \textbf{const-q} model is given the correct $q$. The solid lines show means and shades with same colors show standard deviation.}
 \label{fig:const-b}
\end{figure}

\begin{figure}[h]
 \centering
 \includegraphics[width=\linewidth]{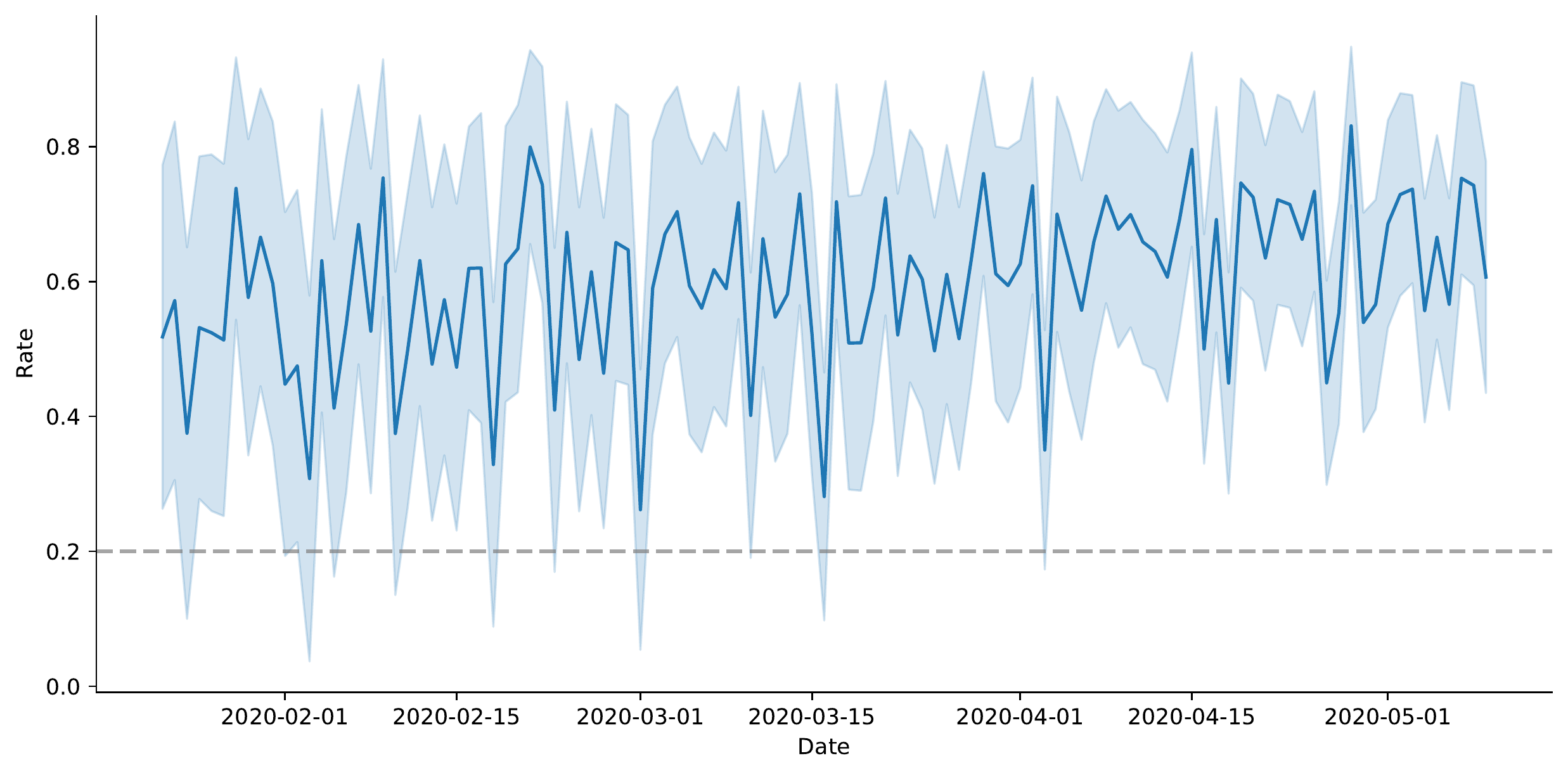}
 \caption{Estimated $q$ for data generated with constant $\beta$ and $q$. The solid lines show means and shades with same colors show standard deviation. The dotted line indicates the true $q$. }
 \label{fig:const-q}
\end{figure}

Fig.~\ref{fig:const-b} and Fig.~\ref{fig:const-q} show the estimated $\beta$ and $q$ for synthesized data using $\beta=0.07$ and $q = 0.2$.
Fig~\ref{fig:const-b} also shows the estimated $\beta$ by \textbf{const-q} model.
the true $q = 0.2$ is given to \textbf{const-q} model as data.

\begin{figure}[h]
 \centering
 \includegraphics[width=\linewidth]{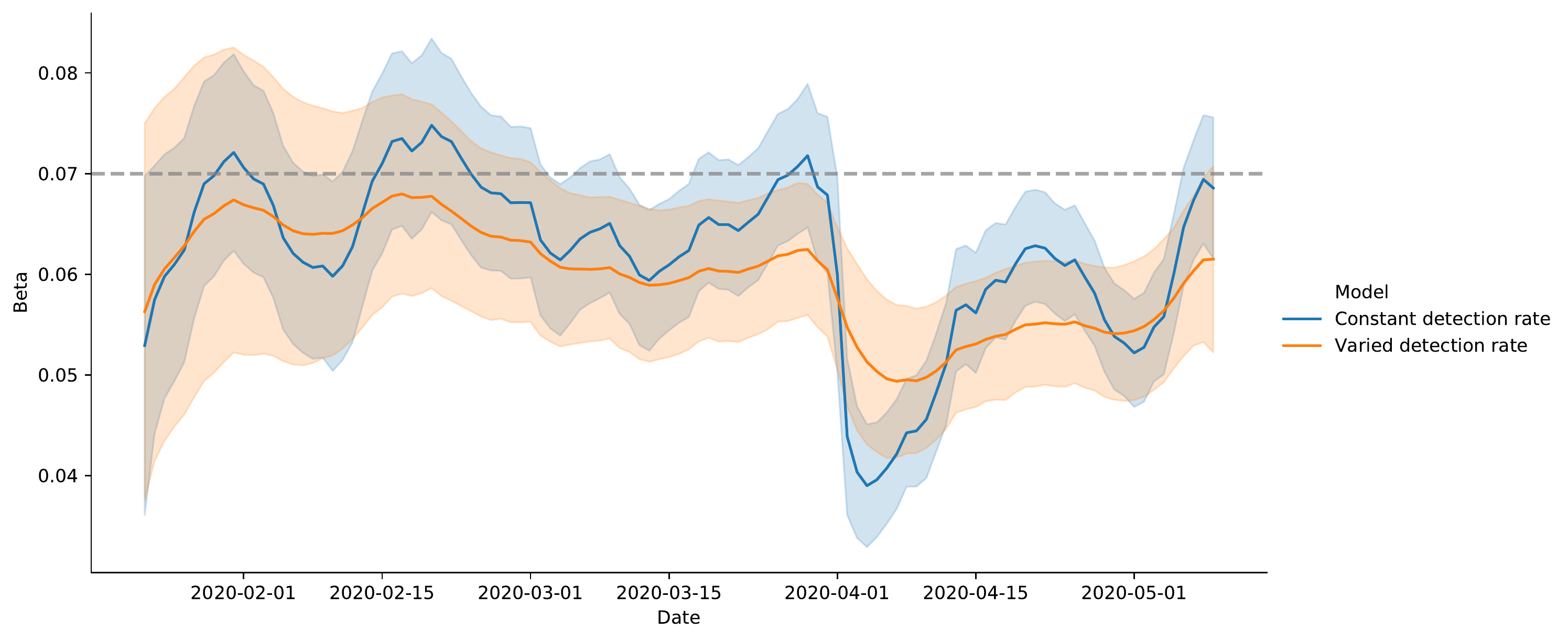}
 \caption{Estimated $\beta$ for data generated with constant $\beta$ and step-like $q$ which starts with $1$ and reduces its value to $0.8$. The blue line indicates the estimate of \textbf{const-q} model while the orange line indicates the estimate of our model. The solid lines show means and shades with same colors show standard deviation. The dotted line indicates the true $\beta$. \textbf{const-q} model is given $q=1$.}
 \label{fig:step-b}
\end{figure}

\begin{figure}[h]
 \centering
 \includegraphics[width=\linewidth]{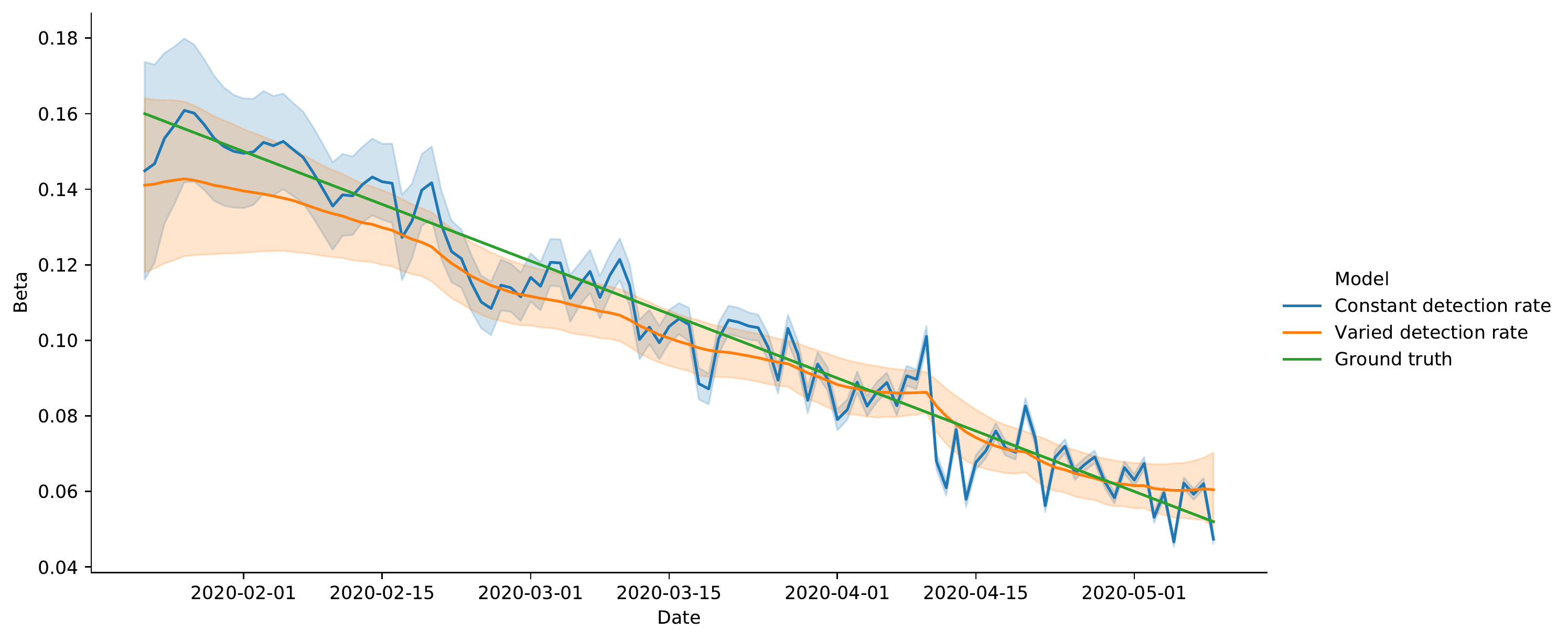}
 \caption{Estimated $\beta$ for data generated with linearly decreasing $\beta$ and noisy $q$. The blue line indicates the estimate of \textbf{const-q} model while the orange line indicates the estimate of our model. The solid lines show means and shades with same colors show standard deviation. The green line indicates the true $\beta$. \textbf{const-q} model is given $q=0.5$.}
 \label{fig:trend-b}
\end{figure}

The results show that both models can estimate $beta$ correctly, while completely fail to estimate $q$.
Estimated $q$ is biased to 1 and noisy.
Therefore, our method cannot estimate the absolute level of the detection rate.
However, the estimate of $\beta$ is still accurate so our method can be used to estimate $\beta$.
In fact, Fig.~\ref{fig:step-b} and Fig.~\ref{fig:trend-b} shows that our model is robust against changing $q$.

\begin{figure}[h]
 \centering
 \includegraphics[width=\linewidth]{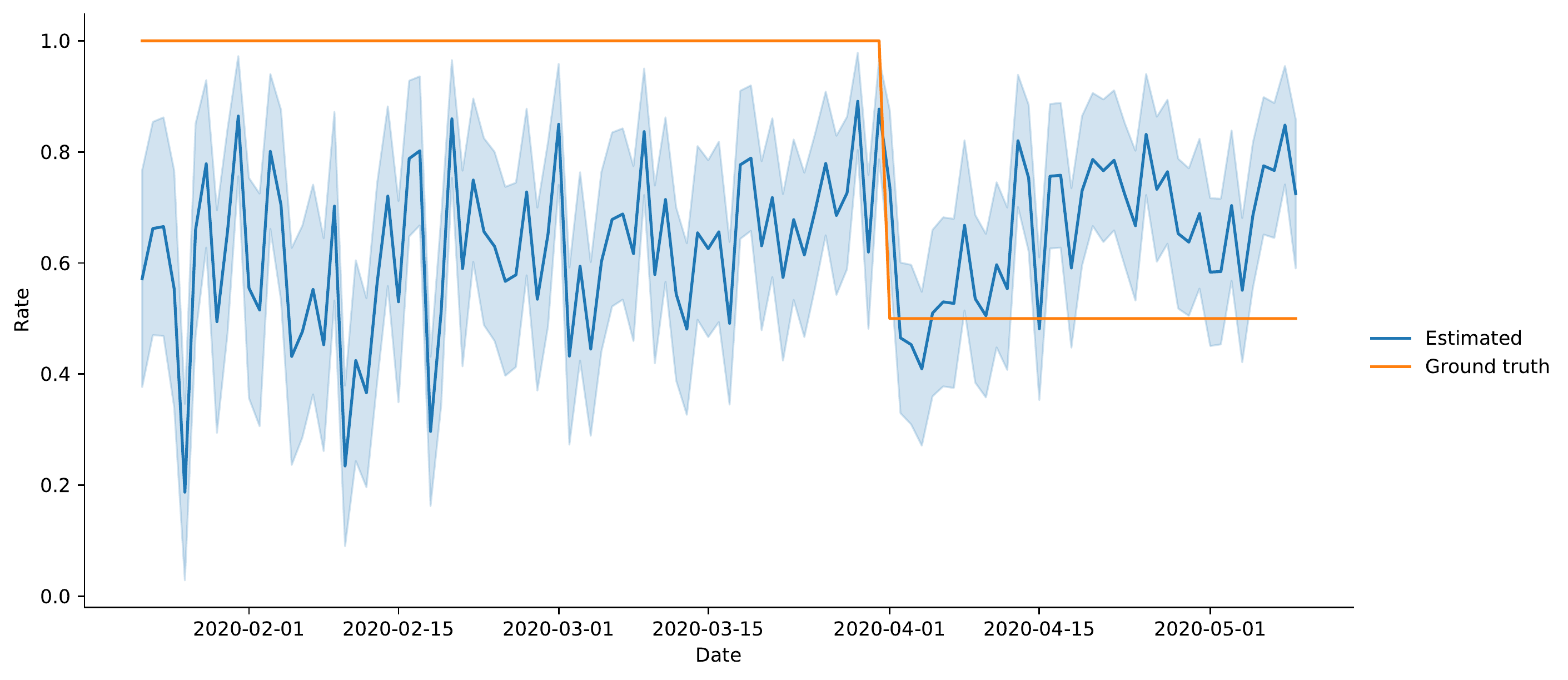}
 \caption{Estimated $q$ for data generated with constant $\beta$ and step-like $q$. The solid lines show means and shades with same colors show standard deviation. The orange line indicates the true $q$.}
 \label{fig:step-q}
\end{figure}

Fig.~\ref{fig:step-b} and Fig.~\ref{fig:step-q} show the estimated $\beta$ and $q$ for synthesized data using constant $\beta=0.07$ but changing $q$ which goes 1 to 0.8 in Apr. 1.

Fig~\ref{fig:step-b} shows $\beta$ estimated by \textbf{const-q} model drops at April 1.
Although the estimate by our model also drops around the same date, the drop is less significant.
This indicates that allowing $q$ vary makes an estimated $\beta$ robust against a sudden change of $q$.

\begin{figure}[h]
 \centering
 \includegraphics[width=\linewidth]{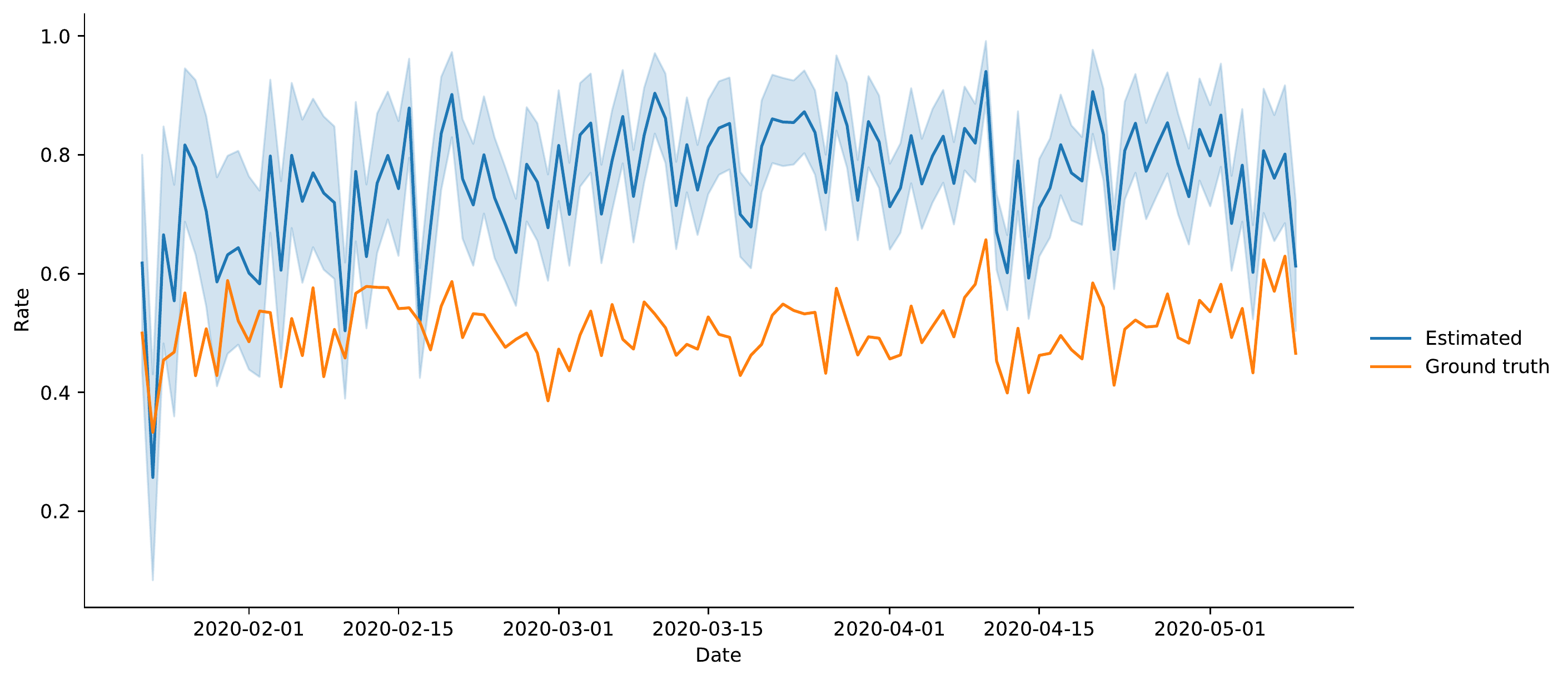}
 \caption{Estimated $q$ for data generated with constant $\beta$ and noisy $q$. The solid lines show means and shades with same colors show standard deviation. The orange line indicates the true $q$.}
 \label{fig:trend-q}
\end{figure}

Fig.~\ref{fig:trend-b} and Fig~\ref{fig:trend-q} show the estimated $\beta$ and $q$ for synthesized data using linearly decreasing $\beta$ and noisy $q$.
Fig.~\ref{fig:trend-b} shows that our method is smooth while the estimate of \textbf{const-q} is noisy.
Again, this shows that our model is more robust against changing $q$.
Further, Fig~\ref{fig:trend-q} indicates that while the estimated $q$ is biased toward 1, the short-term change coincides the ground truth.
Therefore, the estimated $q$ is still useful to find a short-term shock to the detection rate.

\begin{figure}[h]
 \centering
 \includegraphics[width=\linewidth]{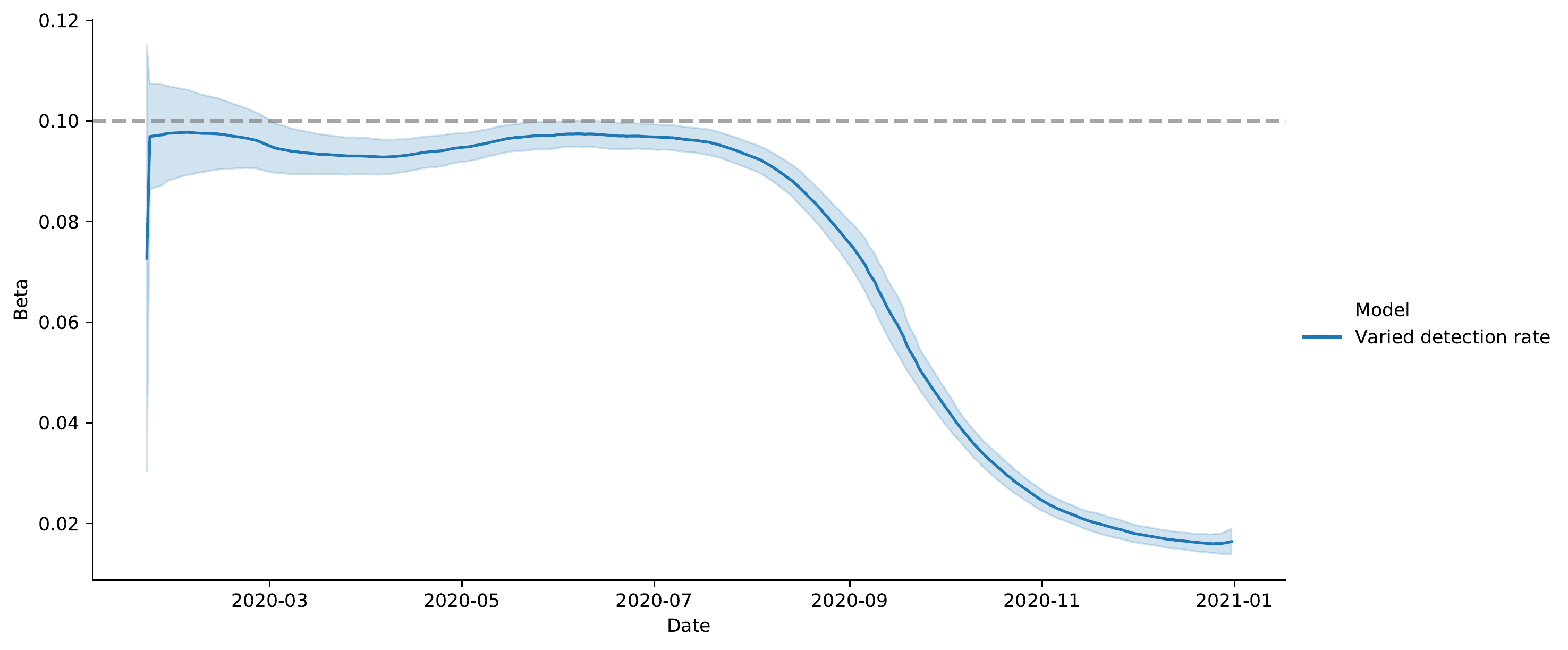}
 \caption{Estimated $\beta$ for data generated with constant $\beta$ and $q$ which make most of population is infected. The blue line indicates the estimate of our model. The solid lines show means and shades with same colors show standard deviation. The dotted line indicates the true $\beta$.}
 \label{fig:saturate-b}
\end{figure}

\begin{figure}[h]
 \centering
 \includegraphics[width=\linewidth]{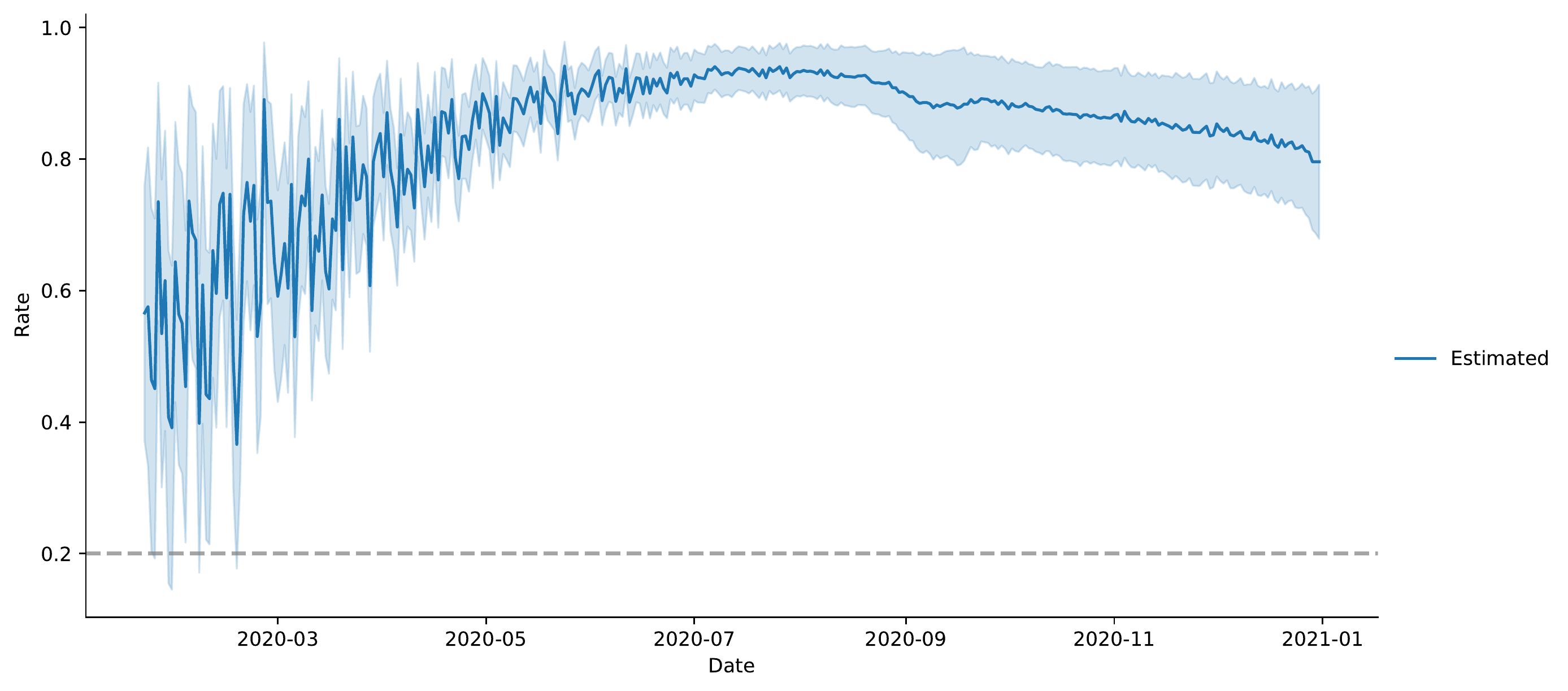}
 \caption{Estimated $q$ for data generated with constant $\beta$ and $q$ which make most of population is infected. The solid lines show means and shades with same colors show standard deviation. The dotted line indicates the true $q$.}
 \label{fig:saturate-q}
\end{figure}

Fig.~\ref{fig:saturate-b} and Fig.~\ref{fig:saturate-q} show the estimate of $\beta$ and $q$ for data generated with constant $\beta$ and $q$.
Unlike Fig.~\ref{fig:const-b} and Fig.~\ref{fig:const-q}, $\beta$ and time horizon are chosen so that most of population are infected in the end.
The result shows that the estimate of $\beta$ becomes unreliable when the infection starts saturated.
Therefore, our model is only useful for the case of a low infection rate.
The estimated $\beta$ by \textbf{const-q} swings widely, therefore is omitted.

In summary, the estimate of $\beta$, which is a key parameter for estimate $R$, by our model is robust against sudden changes and noise in $q$, even though the estimate of $q$ itself is unreliable.
When the infection begins to saturate among population, the estimate of $\beta$ as well as $q$ becomes unreliable.

\subsection{Analysis of Japan}

Next, we present our analysis of Japanese data.

First, we applied three models, \textbf{const}, \textbf{const-q} and our model (\textbf{varied-q}) to Japanese data.
For all parameters of all models, $\hat{R} < 1.1$ holds so convergence is excellent.
Then two information criteria, PSIS-LOO-CV and WAIC are applied to \textbf{const}, \textbf{const-q} and \textbf{varied-q}.
The results are shown in Table~\ref{tbl:IC}.

\begin{table}[h]
 \begin{center}
 \begin{tabular}{lrrrr} \toprule
 Model & PSIS-LOO & DSE(PSIS-LOO) & WAIC & DSE(WAIC) \\ \midrule 
 \textbf{varied-q} & 792.209 & 0 & 720.258 & 0 \\
 \textbf{const-q} & 809.442 & 6.44946 & 730.481 & 3.76378\\
 \textbf{const} & 5056.46 & 809.672 & 5768.74 & 1042.7\\ 
 \bottomrule\\
 \end{tabular}
 \caption{Comparison of PSIS-LOO and WAIC for each model. \textbf{varied-q} is a model with varying $\beta$ and $q$. \textbf{const-q} is a model with varying $\beta$ and fixed $q$. \textbf{const} is a model with fixed $\beta$ and $q$.
 DSE is the standard error of the distance from the best model.}
 \label{tbl:IC}
 \end{center}
\end{table}

Table~\ref{tbl:IC} shows that our model, \textbf{varied-q}, is the best model.
However, we need to be careful because the difference between \textbf{const-q} is not large.
Further, there is a data point in which the Pareto-k for the importance weight distribution is larger than 0.7.
Therefore, the model is not robust and influenced too much from small numbers of data points.

Keeping this in mind, we tentatively choose \textbf{varied-q} and we analyze the result of \textbf{varied-q}.

\begin{figure}[h]
 \centering
 \includegraphics[width=\linewidth]{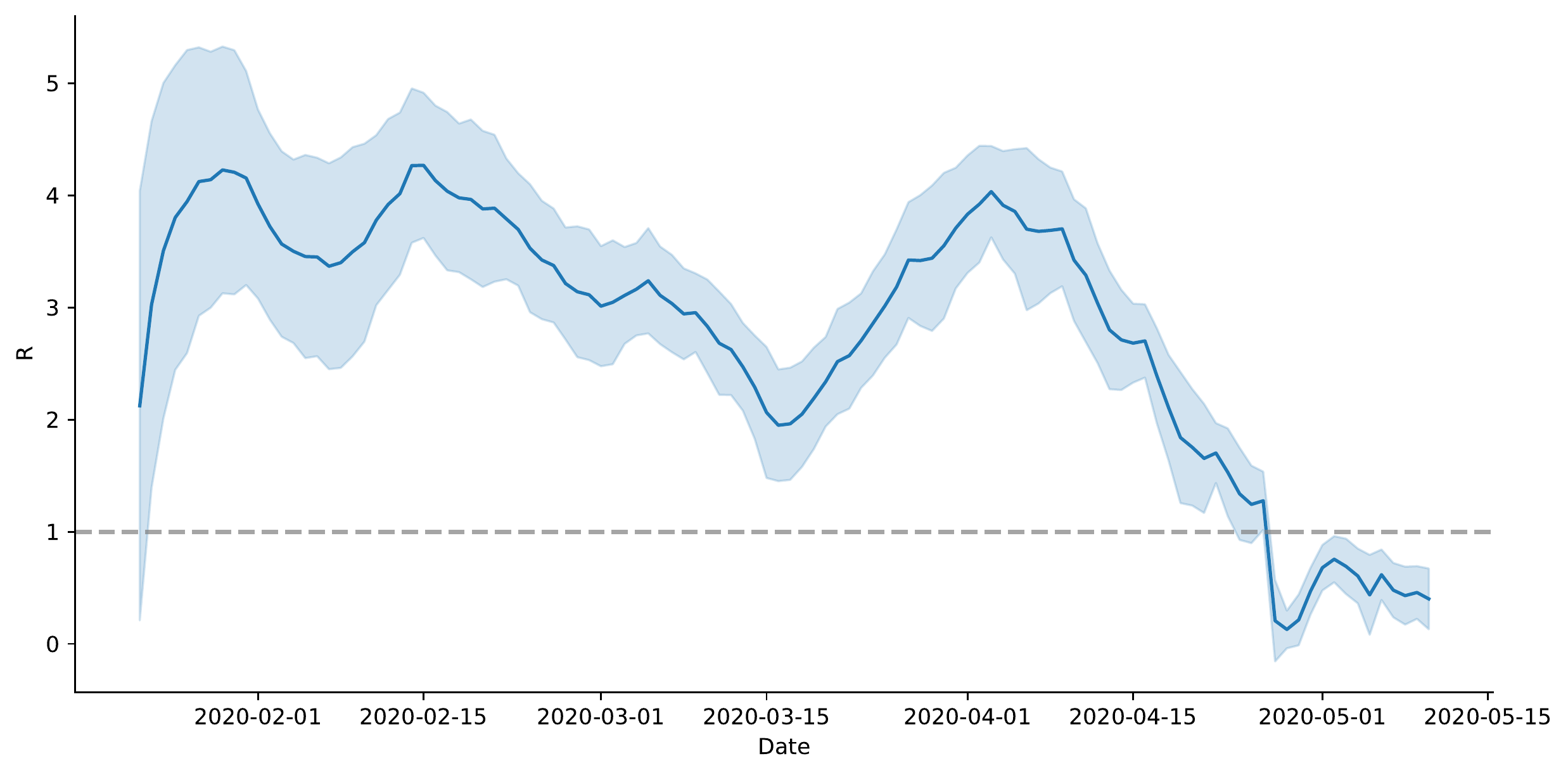}
 \caption{Daily change of $R$ in Japan. The solid lines show means and shades with same colors show standard deviation.}
 \label{fig:b}
\end{figure}

Fig.~\ref{fig:b} shows an estimated $R$ for each day in Japan by \textbf{varied-q}.
The first upward trend may not be very reliable because there is a few data for infection.
The downward trend from the mid. February until the mid. March could be explained by public awareness and tracking effort of infection.
After the mid. March, the tracking effort might be overwhelmed, thus created an upward trend until the beginning of April, when ``the state of emergency'' was declared to major urban areas.
Since then, there was a downward trend, and now $R$ is below $1$.
Thus, currently the infection is shrinking.

\begin{figure}[h]
 \centering
 \includegraphics[width=\linewidth]{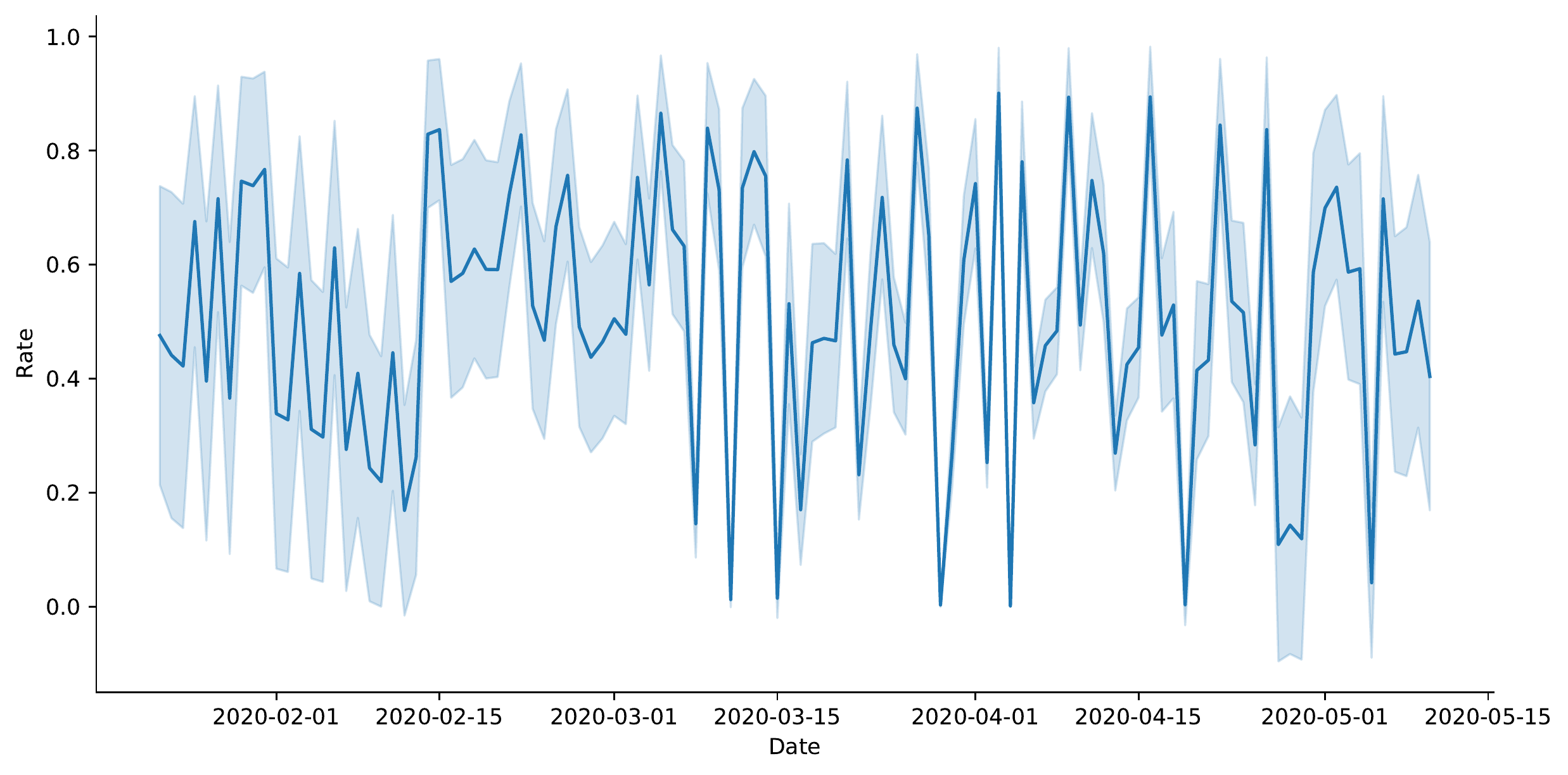}
 \caption{Daily change of the detection rate in Japan. The solid lines show means and shades with same colors show standard deviation.}
 \label{fig:q}
\end{figure}

\subsection{Sensitivity analysis}

Before going analysis of other countries, we analyze sensitivity of the result to prior distributions using Japanese data as an example.

Fig.~\ref{fig:b} shows an estimated $R$ for each day in Japan by \textbf{varied-q}.
\begin{figure}[h]
 \centering
 \includegraphics[width=\linewidth]{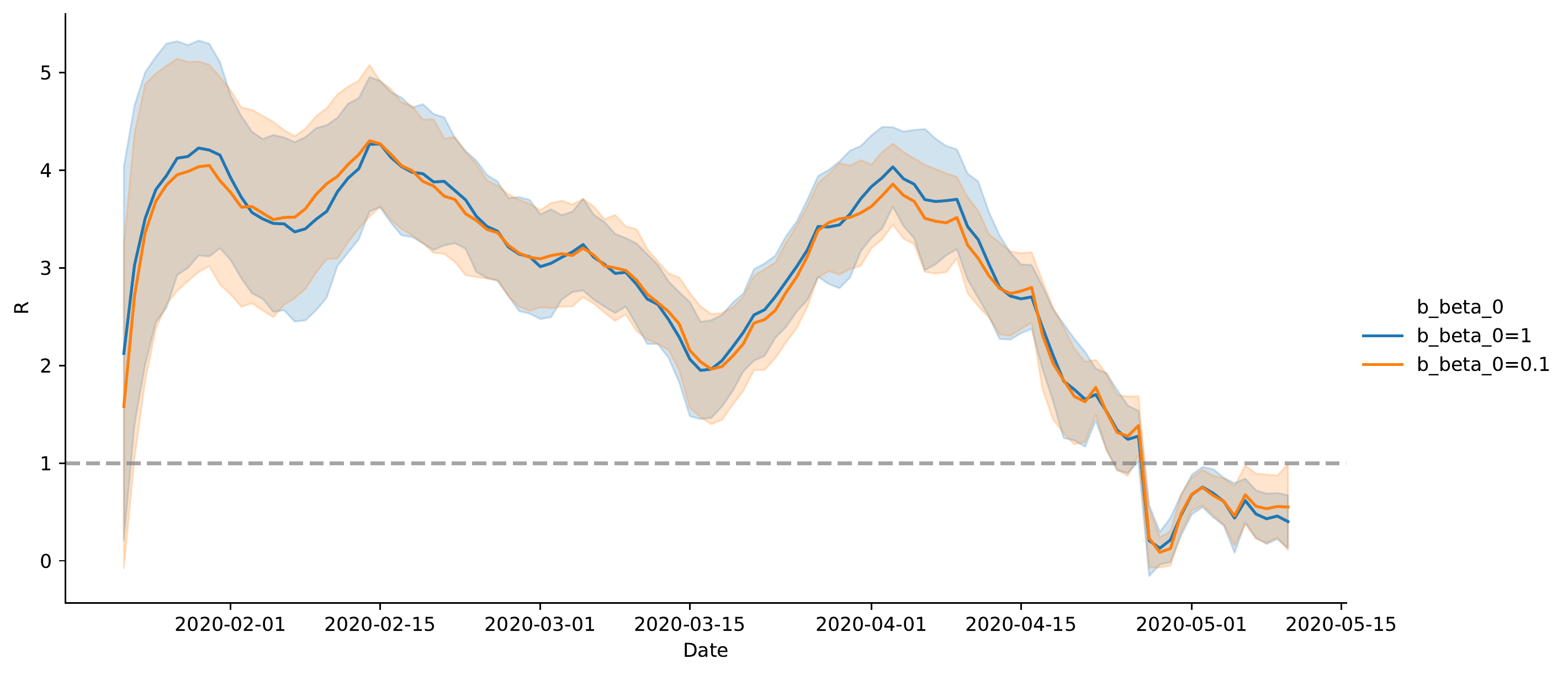}
 \caption{Daily estimate of $R$ in Japan using different priors. The solid lines show means and shades with same colors show standard deviation.}
 \label{fig:sensitivity}
\end{figure}
The blue line shows the estimate of $R$ based on the prior $\sigma_b \sim \textup{Exponential}(1)$, which are used throughout in this paper.
The orange line shows the estimate of $R$ based on the prior $\sigma_b \sim \textup{Exponential}(0.1)$.
There is no significant difference of estimates between two priors; therefore we can safely conclude that the effect of prior is small.

\subsection{Comparison of Denmark and Sweden}

Next, we compare Denmark and Sweden.
These countries are economically and socially similar countries, but employed very different policies against COVID-19.

We applied the same procedure as Japan to data from two countries.
All parameters are converged and information criteria favor \textbf{varied-q} model.
\begin{figure}[h]
 \centering
 \includegraphics[width=\linewidth]{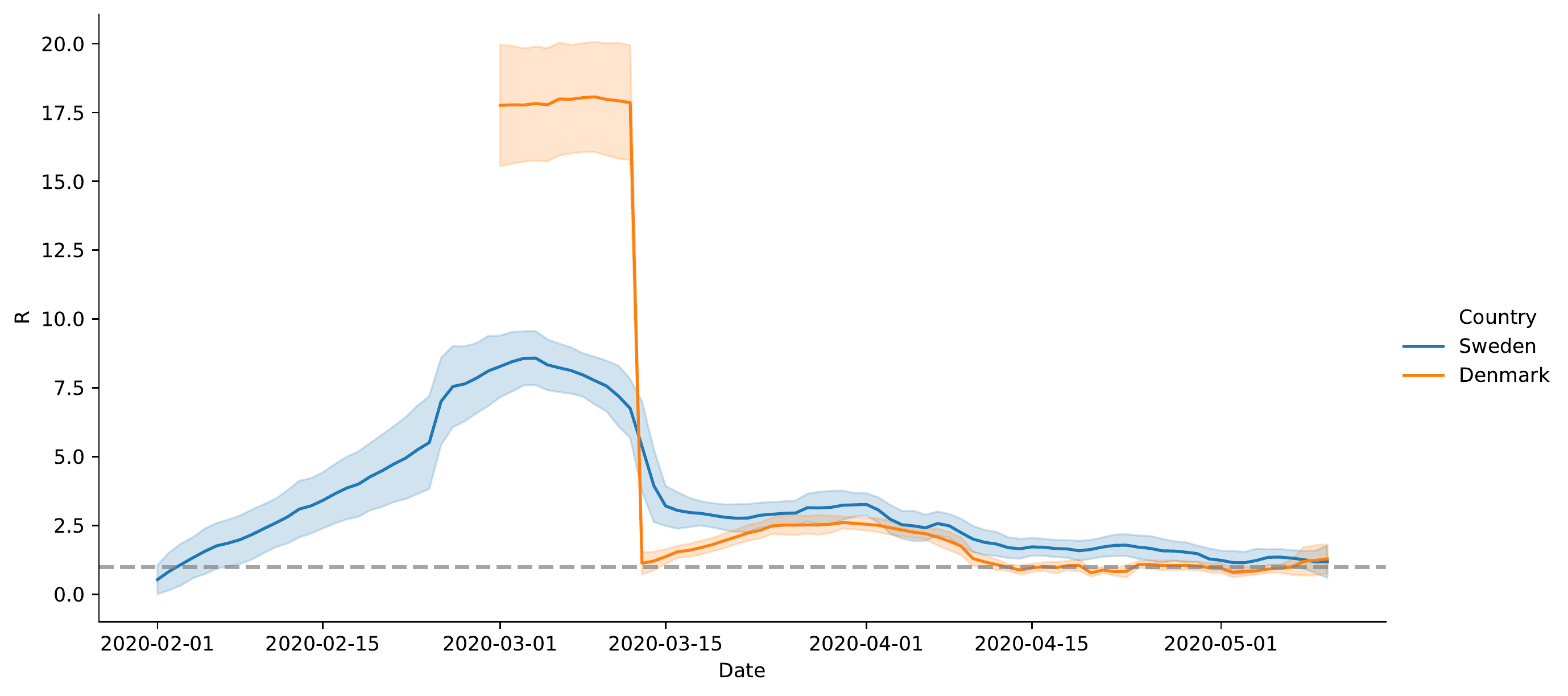}
 \caption{Daily estimate of $R$ in Denmark and Sweden. The solid lines show means and shades with same colors show standard deviation.}
 \label{fig:nordic}
\end{figure}
For Sweden, we only applied data from Mar 1. because the confirmed cases only appear in February 27.
The estimated $R$ clearly shows that lock-down introduced March 13. was effective to put down $R$.
However, in Sweden, which did not employ lock-down, $R$ reduced gradually and now $R$ is almost same between Denmark and Sweden.
This might suggest that lock-down is not very effective in long run but might also due to unfavorable conditions to Denmark, for example, higher population density.

\subsection{Multi-national comparison}

\begin{figure}[h]
 \centering
 \includegraphics[width=\linewidth]{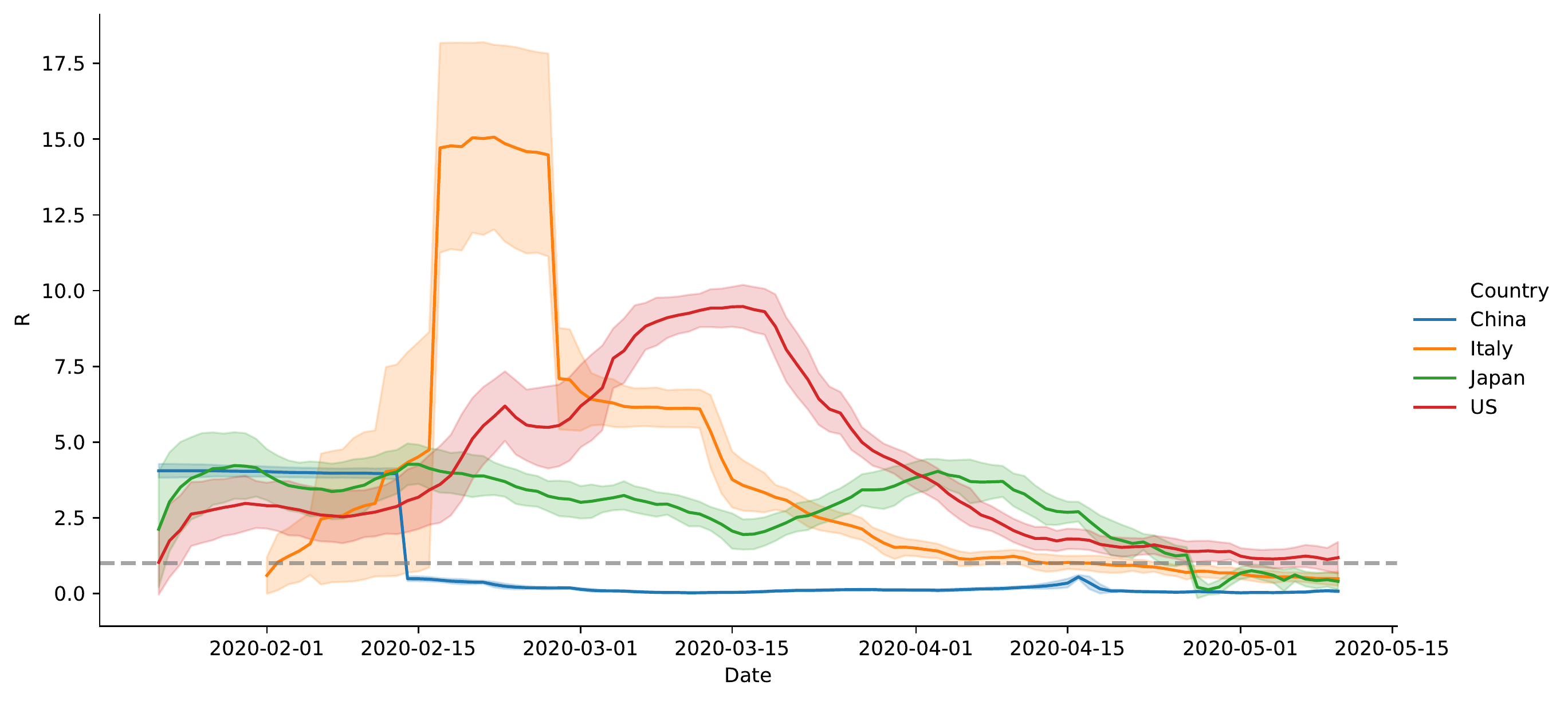}
 \caption{Daily estimate of $R$ in several countries. The solid lines show means and shades with same colors show standard deviation.}
 \label{fig:multi-countries}
\end{figure}

Fig.~\ref{fig:multi-countries} shows daily estimate of $R$ of China, Italy, Japan and US.
The same method as Japan was applied to the rest of countries and adequacy of \textbf{varied-q} model was verified.
For Italy, we only apply data from March 1, 2020 because otherwise the parameter estimation did not converge.
The same method was applied to Korea but the parameter estimation does not converge.

The results show that China, Italy, Japan, and the US are about to exit epidemic.

\section{Future work}

The method used in this paper has several limitations.

First, as experiment used simulation data revealed, our method cannot determine the true level of detection rate, nor its long-term trends.
To find the true detection rate, we need different kind of data, such as excess mortality.

Second, our model uses a naive SIR model and does not consider incubation period and reporting delay.
We can reconstruct the date of exposure by back projection, so it would be interesting apply our method to such data.

\section*{Acknowledgement}

The author thanks to Kentaro Matsuura and the Tokyo.R Slack group for many suggestions and advice on Bayesian modeling.
The author also thanks to Peter Turchin, from whose work the author's work started.

\bibliographystyle{plain}
\bibliography{BayesianCOVID-19}

\end{document}